\documentclass[prl,twocolumn,groupedaddress,showpacs,floatfix]{revtex4}

\usepackage[usenames, dvipsnames]{color}
\usepackage{graphicx}
\usepackage{bm}
\usepackage{amsmath}
\usepackage{amsfonts}
\usepackage{amssymb}
\usepackage{latexsym}
\usepackage{cancel}
\usepackage{bbm}
\usepackage{hyperref}

\newcommand{\Ket}[1]{|#1\rangle}
\newcommand{\Bra}[1]{\langle#1|}
\newcommand{\Ketbra}[2]{|#1\rangle\!\langle#2|}
\newcommand{\Braket}[1]{\langle #1 \rangle}

\begin{document}

\title{Robust Entanglement through Macroscopic Quantum Jumps}
\author{Jeremy Metz, Michael Trupke, and Almut Beige}
\affiliation{Blackett Laboratory, Imperial College London,
Prince Consort Road, London SW7 2BZ, United Kingdom}

\date{\today}

\begin{abstract}
We propose an entanglement generation scheme that requires neither the coherent evolution of a quantum system nor the detection of single photons. Instead, the desired state is heralded by a {\em macroscopic} quantum jump. Macroscopic quantum jumps manifest themselves as a random telegraph signal with long intervals of intense fluorescence (light periods) interrupted by the complete absence of photons (dark periods). Here we show that a system of two atoms  trapped inside an optical cavity can be designed such that a dark period prepares the atoms in a maximally entangled ground state. Achieving fidelities above $0.9$ is possible even when the single-atom cooperativity parameter $C$ is as low as $10$ and when using a photon detector with an efficiency as low as $\eta = 0.2$.
\end{abstract}
\pacs{03.67.Mn, 03.67.Pp, 42.50.Lc }

\maketitle

There are countless applications for highly entangled quantum states,
ranging from the improvement of frequency standards \cite{Huelga} to quantum
information processing, where they can be utilised for one-way quantum
computing \cite{Briegel}. Consequently much effort has been made over the
last few years to generate highly entangled states in the laboratory. For
example, groups in Boulder and Innsbruck have entangled up to eight ions \cite{blatt} and a four-photon cluster state has been created by Walther {\em et al.}~\cite{walther}.
However, scaling these setups to many more qubits is not straightforward.
Adding qubits in ion traps increases the density of motional states and
therefore requires some form of distributed quantum computing, possibly
involving ion transport \cite{kielpinski}. The main difficulties when
entangling photons are the lack of an effective interaction and reliable
photon storage.

Although not yet demonstrated experimentally, a promising alternative is to
entangle atoms by coupling them via an optical cavity. The ability to
position the atoms inside such a resonator to within a fraction of the
optical wavelength \cite{Meschede} together
with the possibility of effectively coupling distant cavities via optical
networks \cite{Lim,Lim2} promises a high degree of control and scalability. The
quality of atom-cavity setups is often measured by the single-atom cooperativity parameter 
$C \equiv g^2/\kappa \Gamma$, which compares the 
atom-cavity coupling constant $g$ with the cavity photon decay rate $\kappa$
\cite{kappa} and the atom decay rate $\Gamma$. After several years of
experimentation with optical cavities values of $C$ of about $50$ have been
reported in the literature \cite{Chapman}. Significantly larger $C$'s
are currently only possible when using AlGaAs resonators with embedded quantum dots
\cite{dots}.

Despite recent progress, the achievable values of $C$ are still too low to allow for the high fidelity generation of highly entangled atomic states with existing proposals. Even when using dissipation-assisted stimulated Raman adiabatic passages (STIRAP) \cite{Pellizzari1995}, strong detunings \cite{ZhengGuo}, or the idea of quantum computing using
dissipation \cite{letter}, a precision of more than $0.85$ requires $C >
100$. An exception is the scheme by Pachos and Walther \cite{PachosWalther}
which achieves better results at the cost of a very slow and complex
entangling STIRAP process. Even probabilistic schemes based on the detection of single photons promise fidelities above $0.9$ only with perfect single photon detectors or require $C \gg1$ \cite{Cabrillo}. In contrast, we show here that it is possible to prepare two atoms in the maximally entangled ground state $|a_{01} \rangle \equiv (|01 \rangle - |10 \rangle)/\sqrt{2}$ with a fidelity $F$ above $0.9$ even if $C$ is as low as $10$
and when using a photon detector with an efficiency as low as $\eta = 0.2$. 
Higher fidelities require a larger value of $C$ or $\eta$.   

\begin{figure}
\begin{minipage}{\columnwidth}
\begin{center}
\resizebox{\columnwidth}{!}{\rotatebox{0}{\includegraphics {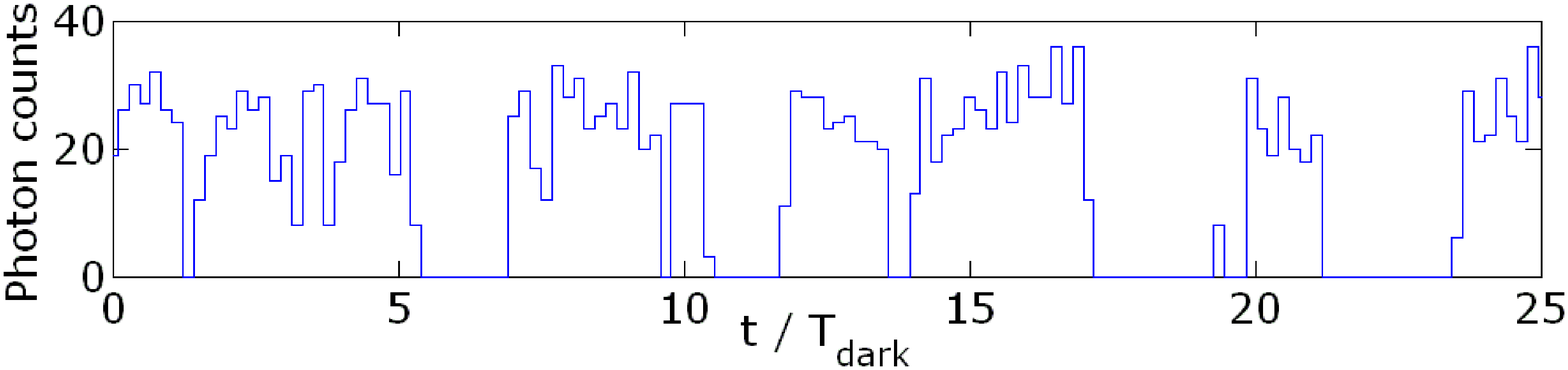}}}
\end{center}
\vspace*{-0.7cm}
\caption{Macroscopic quantum jumps in the fluorescence of two atoms trapped inside an optical cavity (c.f.~Figure \ref{fig:level+setup}) obtained from the numerical simulation of a possible trajectory assuming $\Delta = 50 \, g$,   $\Gamma_0 = \Gamma_1$, $\kappa = \Omega_{\rm L} =  g$ and $\Gamma = \Omega_{\rm M} = 0.1 \, g$, i.e.~$C=10$. Shown is the number of photon emissions within time intervals of length $\Delta t = 0.38 \, T_{\rm dark}$ as a function of time.} \label{fig:toymodel}
\end{minipage}
\end{figure}

Achieving this is only possible, when dissipation plays a major role in the state preparation process. In the following we achieve this by employing a quantum mechanical phenomena known as {\em quantum jumps}. These abrupt transitions of the internal states of an atom, upon the emission or absorption of a light quantum, were proposed by Bohr as early as 1913 \cite{Bohr}. Bohr's quantum jumps were disturbing to many prominent physicists, including Schr\"odinger, as they raised many questions about our understanding of quantum mechanics \cite{Zoller}. However, it became possible to observe these jumps experimentally in the form of {\em macroscopic} light and dark periods, like the ones shown in Figure \ref{fig:toymodel}. In 1986, several group reported the blinking of the fluorescence of a single laser driven trapped ion \cite{Toschek}.

In the language of modern quantum theory, macroscopic quantum jumps are a random telegraph process with long intervals of intense photon emissions interrupted by periods of the complete absence of photons. They occur in the fluorescence of a {\em single} ion, if rapidly repeated measurements project the system either in a subspace of states, in which the systems emits photons at a high rate, or in a state, where it cannot undergo spontaneous emissions \cite{javanainen}. Dehmelt, who predicted the existence of macroscopic quantum jumps in 1975, rightly related this effect to {\em electron shelving} \cite{shelving}. Small deviations from ideal measurements can cause a sudden change of the respective measurement outcome and are thereby responsible for transitions (jumps) from a light into a dark period and vice versa \cite{behe}. 

In the following, we generalise the setup introduced by Dehmelt and propose a scheme to entangle the electronic ground states of two atoms trapped inside an optical resonator. Interactions are applied such that the combined atom-cavity system generates macroscopic light and dark periods. Moreover, the desired state is the only dark state such that the absence of fluorescence indicates the shelving of the system with a high precision into the desired state. Turning off the applied laser fields within a dark period stops the system from returning into yet another light period and is enough to complete the state preparation.

\begin{figure}
\begin{minipage}{\columnwidth}
\begin{center}
\resizebox{\columnwidth}{!}{\rotatebox{0}{\includegraphics {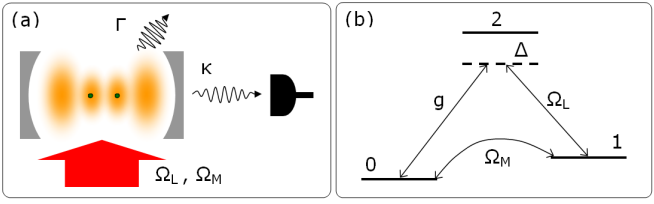}}}
\end{center}
\vspace*{-0.5cm}
\caption{(a) Experimental setup for the preparation of a maximally entangled state of two atoms trapped inside an optical cavity. (b) Level configuration of one atom with the qubit states $|0 \rangle$ and $|1 \rangle$.}
\label{fig:level+setup}
\end{minipage}
\end{figure}

The concrete experimental setup and the required atomic level diagram are
shown in Figure \ref{fig:level+setup}. The trapping of the particles and the
directions of the incoming laser fields should be chosen such that both atoms
experience the same coupling constants. More concretely, the 0--2 transition
of each atom couples to the cavity field with coupling strength $g$, while
laser fields with Rabi frequency $\Omega_{\rm M}$ and $\Omega_{\rm L}$ drive
the 0--1 and the 1--2 transitions, respectively. Here we are interested in
the parameter regime where \cite{note}
\begin{equation} \label{par}
\Omega_{\rm M} < g , \, \kappa , \, \Gamma  , \, \Omega_{\rm L} \ll \Delta  \, ,
\end{equation}
with the detuning $\Delta $ as shown in Figure \ref{fig:level+setup} and
where $\Gamma$ denotes the spontaneous decay rate of level 2. $\Omega_{\rm
M}$ can be realised using a microwave, a two-photon Raman transition
via level 2 or a fourth level, since selection rules forbid the direct
excitation of the 0--1 transition.

In the following we model the atom-cavity system shown in Figure
\ref{fig:level+setup} by the master equation
\begin{equation} \label{master}
\dot \rho = - {\textstyle {{\rm i} \over \hbar}} \, \big[ H_{\rm cond} \rho
- \rho H_{\rm cond}^\dagger \big] + {\cal R} (\rho) \, .
\end{equation}
Here ${\cal R}$ is the reset operator describing the change of the system in
the event of an emission and $H_{\rm cond}$ is the conditional Hamiltonian
relating to the no-photon evolution \cite{behe}. If $b$ is the
annihilation operator for a photon inside the resonator, then $H_{\rm cond}$
equals
\begin{eqnarray} \label{Hcond}
H_{\rm cond} &=& \sum_{i=1,2} \big[ \, {\textstyle {1 \over 2}} \hbar \Omega_{\rm L} \, \Ket{1}_{ii}\Bra{2} + {\textstyle {1 \over 2}}  \hbar \Omega_{\rm M} \, \Ket{0}_{ii}\Bra{1} + {\rm H.c.} \nonumber\\
&& + \hbar g \, \Ket{0}_{ii}\Bra{2} \, b^\dagger + {\rm H.c.} + \hbar \big( \Delta - {\textstyle {{\rm i} \over 2}} \Gamma \big) \, |2 \rangle_{ii} \langle 2| \, \big] \nonumber\\
&& - {\textstyle {{\rm i} \over 2}} \hbar \kappa \, b^\dagger b
\end{eqnarray}
in the interaction picture with respect to the interaction-free Hamiltonian
and within the rotating wave approximation. Moreover ${\cal R}(\rho)$ is
given by
\begin{eqnarray} \label{RRR}
{\cal R}(\rho) &=& \sum_{j=0,1} \sum_{i=1,2} \Gamma_j \, |j \rangle_{ii} \langle 2| \, \rho \, |2 \rangle_{ii} \langle j| + \kappa \, b \rho b^\dagger
\end{eqnarray}
with $\Gamma_j$ being the spontaneous decay rate of the 2--$j$ transition
$(\Gamma = \Gamma_0 + \Gamma_1)$. 

As the operators (\ref{Hcond}) and (\ref{RRR}) treat both atoms in exactly the same way, it is convenient to
introduce the states $\Ket{s_{jk}} \equiv (\Ket{jk} + \Ket{kj})/\sqrt{2}$
and $\Ket{a_{jk}} \equiv (\Ket{jk} - \Ket{kj})/\sqrt{2}$. Using this
notation, Eq.~(\ref{Hcond}) becomes
\begin{widetext}
\begin{eqnarray} \label{Hcond2}
H_{\rm cond} &=& {\textstyle {1 \over 2}} \hbar \Omega_{\rm L} \big[ \Ket{s_{01}}\Bra{s_{02}} +  \Ket{a_{01}}\Bra{a_{02}} + \sqrt{2} \big( \Ket{11}\Bra{s_{12}} + \Ket{s_{12}}\Bra{22} \big) + {\rm H.c.} \big] +  {\textstyle {1 \over 2}} \hbar \Omega_{\rm M} \big[ \Ket{s_{02}}\Bra{s_{12}}  +  \Ket{a_{02}}\Bra{a_{12}} \nonumber \\
&& + \sqrt{2} \big( \Ket{00}\Bra{s_{01}} + \Ket{s_{01}}\Bra{11} \big) + {\rm H.c.} \big] + \hbar g \big[
\Ket{s_{01}}\Bra{s_{12}} b^\dagger - \Ket{a_{01}}\Bra{a_{12}} b^\dagger + \sqrt{2} \big( \Ket{00}\Bra{s_{02}} + \Ket{s_{02}}\Bra{22} \big) b^\dagger + {\rm H.c.} \big] \nonumber\\
&& - {\textstyle {{\rm i} \over 2}} \hbar \kappa \, b^\dagger b + \hbar \big( \Delta - {\textstyle {{\rm i} \over 2}} \Gamma \big) \big[ \Ket{s_{02}}\Bra{s_{02}} +  \Ket{a_{02}}\Bra{a_{02}} + \Ket{s_{12}}\Bra{s_{12}} + \Ket{a_{12}}\Bra{a_{12}} + 2 \Ket{22}\Bra{22} \big]  \, .
\end{eqnarray}
\end{widetext}
Similarly, Eq.~(\ref{RRR}) can be written as
\begin{eqnarray} \label{jump}
{\cal R}(\rho) &=& \sum_{j=0,1} \sum_{i=1,2} \Gamma_j \, R_{ji} \, \rho \, R_{ji}^\dagger +  \kappa b \rho \, b^{\dag}
\end{eqnarray}
with the reset operators
\begin{eqnarray} \label{jump3}
R_{01} &\equiv& \Ketbra{00}{s_{02}} + {\textstyle {1 \over \sqrt{2}}} \big( \Ketbra{s_{01}}{s_{12}} - \Ketbra{a_{01}}{a_{12}} \big) + \Ketbra{s_{02}}{22} \, , \nonumber\\
R_{02} &\equiv& \Ketbra{00}{a_{02}} + {\textstyle {1 \over \sqrt{2}}} \big( \Ketbra{s_{01}}{a_{12}} - \Ketbra{a_{01}}{s_{12}} \big) - \Ketbra{a_{02}}{22} \, , \nonumber\\
R_{11} &\equiv& \Ketbra{11}{s_{12}} + {\textstyle {1 \over \sqrt{2}}} \big( \Ketbra{s_{01}}{s_{02}} + \Ketbra{a_{01}}{a_{02}} \big) + \Ketbra{s_{12}}{22} \, , \nonumber \\
R_{12} &\equiv& \Ketbra{11}{a_{12}} + {\textstyle {1 \over \sqrt{2}}} \big( \Ketbra{s_{01}}{a_{02}} + \Ketbra{a_{01}}{s_{02}} \big) - \Ketbra{a_{12}}{22} \, . \nonumber \\
\end{eqnarray}

\begin{figure}
\begin{minipage}{\columnwidth}
\begin{center}
\resizebox{\columnwidth}{!}{\rotatebox{0}{\includegraphics {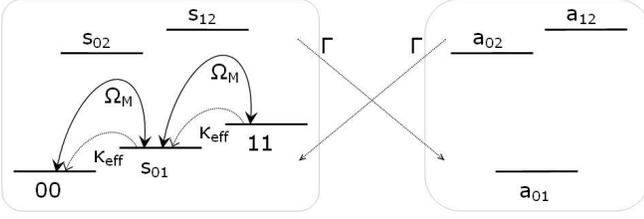}}}
\end{center}
\vspace*{-0.5cm}
\caption{Effective evolution within a light and a dark period involving
the states $|00 \rangle$, $|s_{01} \rangle$, $|11 \rangle$ and $|a_{01}
\rangle$, respectively. Transitions between the subspaces can occur after an
emission from the excited atomic states $|s_{02} \rangle$, $|s_{12} \rangle$
and $|a_{02} \rangle$.} \label{fig:delta_like_toy}
\end{minipage}
\end{figure}

Given the parameter regime (\ref{par}), the excited atomic states with population in $|2 \rangle$ evolve much faster than all other states and can be adiabatically eliminated. Moreover, coherent transitions between states with different numbers of photons in the cavity take place with the frequency $g_{\rm eff} \equiv - \Omega_{\rm L} g/\big(\sqrt{2} \Delta \big) \ll \kappa$. We can therefore also eliminate the states with more than one photon in the resonator. Doing so and denoting the number of cavity photons by $n$, while $\alpha_{jk,n}$, $\sigma_{jk,n}$ and $\xi_{jj,n}$ are the amplitudes of the states $|a_{jk},n \rangle$, $|s_{jk},n \rangle$ and $|jj,n \rangle$, respectively, we find that
\begin{eqnarray} \label{pop}
&& \hspace*{-0.4cm} \alpha_{02,0} = - (\Omega_{\rm L} /2 \Delta) \,  \alpha_{01,0} \, , ~ 
\alpha_{12,0} = \xi_{22,0} = 0  \, , \nonumber \\
&& \hspace*{-0.4cm} \sigma_{02,0} = - (\Omega_{\rm L} /2 \Delta) \, \sigma_{01,0}  \, , ~ 
\sigma_{12,0} = (\Omega_{\rm L} /\sqrt{2} \Delta) \, \xi_{11,0} \, , ~ \nonumber \\ 
&& \hspace*{-0.4cm} \xi_{00,1} = - (2 {\rm i} g_{\rm eff} /\kappa) \, \sigma_{01,0} \, , ~ 
\sigma_{01,1} = - (2 {\rm i} g_{\rm eff} /\kappa) \, \xi_{11,0} ~~~~
\end{eqnarray}
up to first order in $1/\Delta$. Effectively the system can be described by 
\begin{eqnarray} \label{master2}
H_{\rm cond} &=& {\textstyle {1 \over \sqrt{2}}} \hbar \Omega_{\rm M} \big[ \, \Ket{00,0}\Bra{s_{01},0} + \Ket{s_{01},0}\Bra{11,0} + {\rm H.c.} \, \big] \nonumber \\
&& + \hbar \Delta_{\rm L} \, \big[ \, \Ket{11,0}\Bra{11,0} - \Ket{00,0}\Bra{00,0} \, \big] \nonumber \\
&& - {\textstyle {{\rm i} \over 2}} \hbar \, \kappa_{\rm eff} \, \big[ \, \Ket{s_{01},0}\Bra{s_{01},0} + \Ket{11,0}\Bra{11,0} \, \big] \, , \nonumber \\
{\cal R}(\rho) &=& \kappa_{\rm eff} \, \big[ \, |00,0 \rangle \langle s_{01},0| + |s_{01},0 \rangle \langle 11,0 | \, \big] \, ,
\end{eqnarray}
with $\Delta_{\rm L} \equiv - \Omega_{\rm L}^2/\big(4 \Delta \big)$ and $\kappa_{\rm eff} \equiv 4 g_{\rm eff}^2/\kappa$.

The {\em dark} state of a system is the state with a negligible spontaneous decay rate. Furthermore, the system's evolution should not be able to transfer this state into one that can cause an emission. From Eqs.~(\ref{master2}), we see that this applies only to the zero eigenstate $|a_{01},0 \rangle$ of $H_{\rm cond}$. From the above calculations we also see that a {\em light} period mainly involves the states $|00,0 \rangle$, $|s_{01},0
\rangle$ and $|11,0 \rangle$. An interaction drives these states with
frequency $\Omega_{\rm M}$ and detuning $\Delta_{\rm L}$ continuously, as
shown in Figure \ref{fig:delta_like_toy}. Population in $|s_{01},0 \rangle$
and $|11,0 \rangle$ can cause a photon to leak out through the cavity
mirrors with the effective rate $\kappa_{\rm eff}$. After a certain time,
these processes result in a stationary state. Using Eq.~(\ref{master2}) and setting $\dot
\rho = 0$ we find the steady state populations
\begin{eqnarray} \label{P}
P_{00, 0} &=& 1 - P_{s_{01}, 0} - P_{11, 0} \, , ~~  \nonumber \\
P_{s_{01},0} &=& (1 + 8 x^2)/(3 + 16 x^2 + 16 x^4) \, , ~ \nonumber \\
P_{11, 0} &=& 1/(3 + 16 x^2 + 16 x^4)
\end{eqnarray}
with $x \equiv - \Omega_{\rm L}^2/4 \Delta \Omega_{\rm M}$.
From Eq.~(\ref{pop}) we see that there is also a small
population in the states $|00,1 \rangle$ and $|s_{01},1 \rangle$ with one
photon in the cavity. One therefore has $\langle n \rangle = (\kappa_{\rm
eff}/\kappa) \, ( P_{s_{01},0} + P_{11,0} )$ which implies
\begin{equation} \label{eqn:dt}
T_{\rm cav}  = ( 3+4x^2) \cdot {\kappa \Delta^2 \over 4 g^2 \Omega_{\rm L}^2}
\end{equation}
for the mean time between two cavity photon emissions, since $T_{\rm cav}  =
1/\kappa \Braket{n}$.

Transitions between light and dark periods occur since population in the
states $|s_{01},0 \rangle$, $|11,0 \rangle$ and $|a_{01},0 \rangle$ results
in small amounts of population in $|s_{02},0 \rangle$, $|s_{12},0 \rangle$,
and $|a_{02},0 \rangle$, respectively (c.f.~Eq.~(\ref{pop})). This
eventually leads to an atomic decay from level 2 which prepares the setup in
a superposition of a light {\em and} the dark state. Now there are three
possibilities: {\em (i)} After a short time the system emits another photon
via an atomic decay and returns into a neither symmetric nor antisymmetric
state, which brings us back to the initial situation. {\em (ii)} No photon
is emitted for a time, which is long compared to $T_{\rm cav}$, preparing the
atoms in $\Ket{a_{01}}$. {\em (iii)} A photon leaks out through the cavity
mirrors and prepares the setup in a symmetric state
(c.f.~Eq.~(\ref{master2})) with no population in the dark state. In the
latter two cases a transition between a light and a dark period may have
occurred. The reason for the projection into one of the subspaces is that
the observation of cavity leakage reveals information about the system. This
measurement projects the atoms either into a symmetric state or into
$|a_{01} \rangle$.

More concretely, the end of a dark period is caused by a spontaneous
emission from the state $|a_{02},0 \rangle$. Its  population within a dark
period can be calculated using Eq.~(\ref{pop}) with $\alpha_{01,0}=1$.
Moreover we see from Eq.~(\ref{jump3}) that such an emission transfers the
atoms with rate $\Gamma_0$ into the state $|00 \rangle$ and with rate ${1
\over 2} \Gamma_1$ into $|s_{01} \rangle$. Both together imply
\begin{equation} \label{Tdark}
T_{\rm dark} = {1 \over 2 \Gamma_0 + \Gamma_1 } \cdot  {8 \Delta^2 \over \Omega_{\rm L}^2} \, .
\end{equation}
Similarly, taking Eqs.~(\ref{pop}) and (\ref{P}) into account, we see that a
light period ends at any time with the probability density ${1 \over 2}
\Gamma_0 \, P_{s_{02},0} + {1 \over 2} \Gamma_1 \, P_{s_{12},0}$ which
yields
\begin{equation} \label{eqn:TlightOverDt}
T_{\rm light} =  {3 + 16 x^2 + 16x^4 \over 2 \Gamma_0 + (1 + 8 x^2) \Gamma_1} \cdot {8 \Delta^2 \over \Omega_{\rm L}^2}
\end{equation}
for the mean length of a light period.

\begin{figure}
\begin{minipage}{\columnwidth}
\begin{center}
\resizebox{\columnwidth}{!}{\rotatebox{0}{\includegraphics {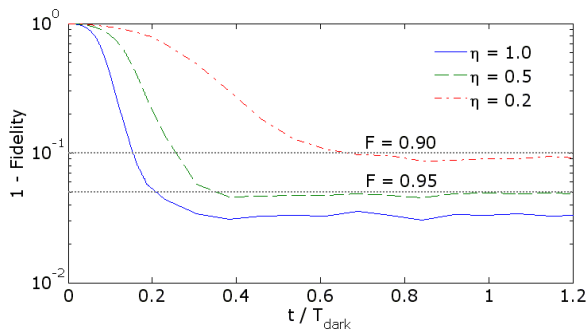}}}
\end{center}
\vspace*{-0.7cm}
\caption{The deviation from unity of the fidelity $F$ of the state prepared
after the {\em detection} of no photon for a time $t$ for different detector
efficiencies $\eta$ and for the same parameters as in Figure \ref{fig:toymodel}
($C=10$) obtained from a numerical simulation.} \label{final}
\end{minipage}
\end{figure}

Crucial for the ability to distinguish a light from a dark period is that
$T_{\rm dark}$ is sufficiently longer than $T_{\rm cav}$. For a wide range
of parameters, namely for $\Omega_{\rm L}^4 \ll (4 \Delta \Omega_{\rm
M})^2$, one has $x^2 \ll 1$ and the ratio $T_{\rm dark}/T_{\rm cav}$ is to a
very good approximation given by its maximum $32 g^2/3\kappa (2 \Gamma_0 +
\Gamma_1)$. Therefore, the obtainable telegraph signal (c.f.~Figure \ref{fig:toymodel})
becomes clearer the larger the single atom cooperativity parameter $C$.
However, the proposed state preparation scheme, namely turning off the laser
fields upon the detection of no photon for a time $t$ of the order of
$T_{\rm dark}$, also works for relatively small values of $C$. For $C=10$ a
dark period is on average $70$ times longer than $T_{\rm cav}$ and the
entangled state $|a_{01} \rangle$ can be obtained with a fidelity above
$0.9$ for $t > 0.7 \, T_{\rm dark}$ even when using a photon detector with
an efficiency of $\eta =0.2$ (c.f.~Figure \ref{final}). Fidelities above
$0.95$ require $\eta \ge 0.5$ and $t > 0.4 \, T_{\rm dark}$. Generally it
does not take long to find the system in a dark period, since $T_{\rm
light}$ is on average only about three times longer than $T_{\rm dark}$.

{\em In summary}, we have described a scheme that can be used to entangle two
atoms trapped inside an optical cavity. The generalisation of our scheme to
a wider class of entangled states, including cluster states for one-way
quantum computing \cite{Briegel}, is straightforward. This can be achieved
by placing more than two atoms simultaneously into the cavity and by moving
them subsequently in and out of the resonator \cite{MePrep}. Problems might
arise from having to trap the atoms at fixed positions with respect to the
cavity mode and the incoming laser fields. However, the techniques for doing
this have improved significantly over recent years
\cite{Meschede}. In addition, the cavity can be used to cool the atomic motion between different attempts to entangle the atoms, as recently demonstrated by Nu{\ss}mann {\em et al.} \cite{cooling}.

Characteristic for the proposed scheme is its constructive use of dissipation. Indeed it is well known that dissipation causes irreversibility and reduces the entropy of coupled quantum systems. Applications are the cooling of single atoms to very low temperatures \cite{cool},  the fusion of Bose Einstein condensates \cite{Jaksch} and measurement-based quantum computing schemes \cite{Briegel,letter,PachosWalther,Lim2}. Moreover, the presence of spontaneous decay rates can be used to stabilise the time evolution of a system \cite{Pellizzari1995}. In contrast to this, the present paper describes a dissipative process, which transfers the system into a {\em macroscopic} dark period. The successful state preparation is heralded by the interruption of a long interval of intense fluorescence. It is therefore possible to entangle atoms with a high fidelity even when using realistic photon detectors and a relatively modest atom-cavity setup. Because of their reliability, we believe that macroscopic quantum jumps will become a very useful tool for the controlled generation of entanglement in physical systems.

\noindent {\em Acknowledgment.} We thank E. Hinds,  P. L. Knight, Y. L. Lim,
D. Meschede's group and A. Serafini for helpful discussions. A. B.
acknowledges support from the Royal Society and the GCHQ. This work was
supported in part by the EU network SCALA and the UK EPSRC through the QIP
IRC.


\begin{thebibliography}{99}
\bibitem{note}
The condition $\Omega_{\rm M} < g , \, \kappa , \, \Gamma  , \, \Omega_{\rm L}$ is not crucial for the scheme to work but simplifies the following calculations.
\end{thebibliography}
\end{document}